\documentclass[         %
aps,                    
prd,                    
showpacs,               
superscriptaddress,    
nofootinbib,            
showkeys,               %
preprintnumbers,        %
floatfix]               
{revtex4}               
\usepackage{graphicx,longtable,amsmath}

\numberwithin{equation}{section}

\begin{document}

\title{Magnetic moments of active and sterile neutrinos}
\author{A.~B. Balantekin}
\email{baha@physics.wisc.edu}
\author{N. Vassh}
\email{vassh@wisc.edu}
\affiliation{Department of Physics, University of Wisconsin, Madison, WI 53706, USA}

\date{\today}

\begin{abstract}
Since most of the neutrino parameters are well-measured, we illustrate precisely the prediction of the Standard Model, minimally extended to allow massive neutrinos, for the electron neutrino magnetic moment. We elaborate on the effects of light sterile neutrinos on the effective electron neutrino magnetic moment measured at the reactors. We explicitly show that the kinematical effects of the neutrino masses are negligible even for light sterile neutrinos. 
\end{abstract}


\maketitle 

\section{Introduction}

Neutrinos primarily interact via weak interactions. Seminal experiments carried out during the last two decades established the long-suspected fact that those interactions couple to particular combinations of mass eigenstates that define the neutrino flavor. Weak interactions preserve the chirality of the particles, so they would be operative even if neutrinos were massless. Neutrinos are electrically neutral particles, but they can have electromagnetic interactions through loops with charged particles. The resulting neutrino dipole moments  would vanish if the neutrino mass were zero since such electromagnetic interactions change the chirality of the particles. However, the same experiments that established the neutrino mixing also established that neutrinos are massive, indicating the presence of neutrino dipole moments. The prediction of the Standard Model, minimally extended to allow massive neutrinos, for the neutrino dipole moments is very small. Now that most of the neutrino parameters are well-measured, we know rather precisely what that prediction is as we elaborate below.  Possible sterile neutrino states do not have the standard weak interaction, but they may have rather large electromagnetic interactions. 

The effects of even a very small neutrino magnetic moment can be amplified in astrophysical settings. For example much of the recent work on the neutrino magnetic moment can be traced back to the hints of the correlation of the solar neutrino flux with solar magnetic activity as either spin precession alone \cite{Okun:1986na} or spin-flavor precession coupled to the matter effects \cite{Lim:1987tk,Minakata:1988gm}. Although evidence for variability of the solar neutrino flux still seems to persist \cite{Sturrock:2008dg}, spin-flavor precession scenario does not play a major role in neutrino propagation in the solar matter \cite{Balantekin:2004tk}. Indeed strict experimental limits on the lack of solar antineutrino flux that would result from spin-flavor precession have been established \cite{Gando:2002ub}. Another example is enhanced neutrino losses due to plasmon decay, 
$\gamma^* \rightarrow \bar{\nu} \nu$, via neutrino dipole moments.  Since neutrinos freely escape the stellar environment this process in turn cools a red giant star faster, delaying helium ignition and increasing the core mass of red giants at the helium flash \cite{Raffelt:1990pj}. The most up-to-date limit is from the red-giant branch in the globular cluster M5: $\mu_{\nu} < 4.5 \times 10^{-12} \mu_B$ (95$\%$ CL) \cite{Viaux:2013hca}. To be precise, the energy-loss argument provides a limit on the sum all neutrino dipole moments and it only applies to neutrinos whose masses do not exceed a few keVs (so that their production is energetically permitted). 

For neutrino-electron scattering the magnetic moment contribution is dominant over the standard electroweak contribution at low recoil energies. Although both solar and reactor neutrinos are used to perform such experiments, the best terrestrial bounds come from reactor experiments. The current best reactor neutrino limit is given by the GEMMA spectrometer at  Kalinin Nuclear Power Plant 
\cite{Beda:2013mta}. GEMMA finds $\mu_{\nu} <  2.9 \times 10^{-11} \mu_B$ at 90\% C.L. with a detector placed at a distance of 13.9 m from the reactor core. Another experiment carried out by the TEXONO collaboration at the Kuo-Sheng Nuclear Power Station finds 
$\mu_{\nu} < 2.2 \times 10^{-10} \mu_B$ with a detector placed at a distance of 28 m from the reactor core \cite{Deniz:2009mu}.

It is possible to interpret various anomalous results from some of the neutrino experiments, astrophysical observations, and cosmology as evidence for the existence of sterile neutrinos \cite{Abazajian:2012ys}. In particular, there may be a discrepancy in short-baseline reactor neutrino experiments between observed and predicted antineutrino fluxes \cite{Mention:2011rk}. This should be treated as a rather tentative conclusion. Although earlier analyses seemed to lend support to the discrepancy (see e.g. \cite{Huber:2011wv}), recent work suggests that nuclear corrections that give rise to this anomaly are very uncertain for the forbidden weak transitions that account for about 30\% of the flux \cite{Hayes:2013wra}. Even though the definitive  solution to this puzzle lies in further experiments \cite{Djurcic:2013oaa}, this reactor anomaly can be interpreted as the signature of additional sterile neutrino states with mass splittings of the order of $\sim 1$ eV$^2$ and oscillation lengths of 3 m. (see e.g. Ref. \cite{Kopp:2013vaa}). If indeed there are light sterile neutrinos that mix with the electron antineutrino then we expect that they would impact neutrino magnetic moment measured at the reactors with detectors placed near the core. 

We briefly review the neutrino magnetic moment measurements at the reactors in the next section. We illustrate precisely the prediction of the Standard Model, minimally extended to allow massive neutrinos, for the neutrino magnetic moment in Section III, where we also discuss the effects of the physics beyond the Standard Model with only three flavors. In Section IV, we consider the case with only one sterile neutrino. We conclude the paper with some brief remarks in Section V. 

\section{Neutrino-electron scattering}

\begin{figure}[b]
\begin{center}
\includegraphics[scale=0.22]{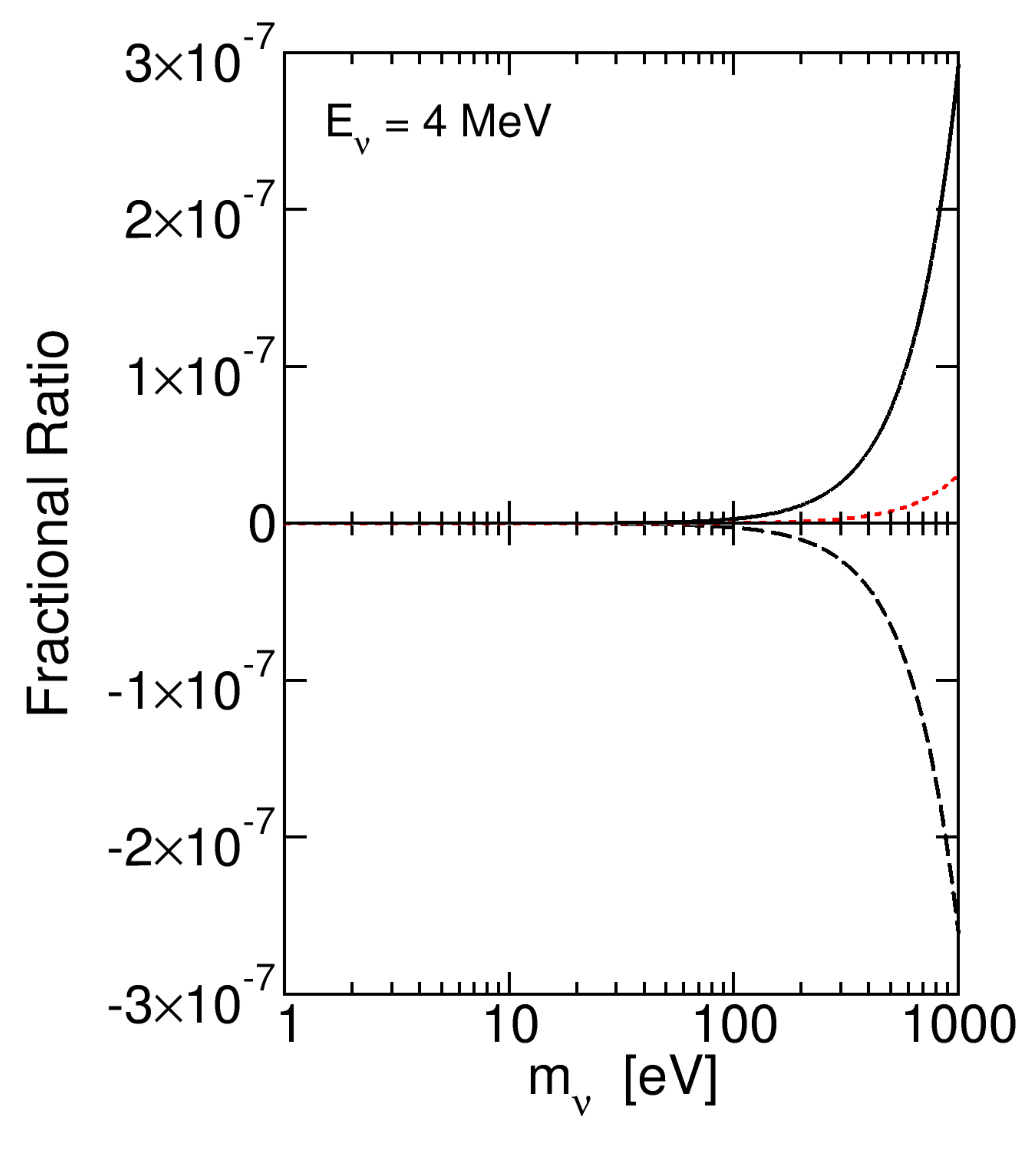}
\caption{(Color online) The fractional ratio of Eq. \ref{fracratio} as a function of the neutrino mass. The solid line is for the case $m_j=0$, i.e. the x-axis of the plot is $m_i$. The dotted line is when $m_i=m_j=$ variable of the x-axis. The dashed line is for the case $m_i=0$, i.e. the variable in the x-axis is $m_j$. The electron recoil energy is taken to be GEMMA's threshold energy of 2.8 keV.}
\label{fig:fig01}
\end{center}
\end{figure}

We first demonstrate that small neutrino masses do not change the kinematics of neutrino-electron scattering via the neutrino magnetic moment. 
To illustrate this consider scattering from neutrino mass eigenstate with mass $m_i$ to the mass eigenstate 
with mass $m_j$. We calculate this cross section to be
\begin{equation}
\label{1}
\frac{d\sigma_{ij}}{dt} = \frac{e^2 \mu_{ij}^2}{8 \pi \lambda} \left[ \frac{1}{t} (2 \lambda + 4 m_e^2 m_i^2 + 2 A \Delta + 2 m_e^2 \Delta + \Delta^2) 
+ (2 A + \Delta) + \frac{2 m_e^2 \Delta^2}{t^2} \right]
\end{equation}
where $\mu_{ij}$ is the neutrino magnetic moment connecting the mass eigenstates $i$ and $j$, $A=s-m_e^2 - m_i^2$, $\Delta = m_i^2 - m_j^2$, and $\lambda = A^2 - 4 m_e^2 m_i^2$. To assess the effect of the finite neutrino masses on the cross section we plot the quantities
\begin{equation}
\left[ \frac{d\sigma_{ij}}{dT_{e}} \left( m_i \neq 0 \>\>\> {\rm and/or} \>\>\> m_j \neq 0 \right) - \frac{d\sigma_{ij}}{dT_{e}} \left( m_i = 0 = m_j  \right) \right] 
\Bigg/
\left[ \frac{d\sigma_{ij}}{dT_{e}} \left( m_i = 0 = m_j \right) \right]
\label{fracratio}
\end{equation}
in Fig. \ref{fig:fig01} for various combinations of neutrino masses. 
In Eq. (\ref{fracratio}), $T_{e}$ is the electron recoil energy. 
Clearly even for unrealistically large neutrino masses, the cross section does not change much and we can safely take neutrino masses to be zero in Eq. (\ref{1}). In the limit $m_i \sim 0 \sim m_j$ Eq. (\ref{1}) reduces to the well-known result 
\cite{Domogatsky:1971tu,Barut:1982fd}: 
\begin{equation}
\label{2}
\frac{d\sigma_{ij}}{dt} = \frac{e^2 \mu_{ij}^2}{4 \pi} \left[ \frac{1}{t} + \frac{1}{s-m_e^2} \right].
\end{equation}
In all the neutrino experiments, the electron antineutrino is in a combination of mass eigenstates. Hence one needs to take into account oscillations between the source and the detector over the distance $L$, leading to an incoherent sum of the individual cross sections \cite{Grimus:1997aa,Beacom:1999wx}:  
\begin{equation}
\label{2a}
\frac{d\sigma}{dt} = \frac{e^2}{4 \pi} \sum_i \left| \sum_j U_{ej} e^{-iE_jL} \mu_{ji}\right|^2
\left[ \frac{1}{t} + \frac{1}{s-m_e^2} \right].
\end{equation}
 
The analysis of the experimental data is usually carried out assuming that electron is initially at rest, i.e., 
\begin{eqnarray}
s &=& m_e^2 + m_i^2 + 2 m_e E_{\nu}, \nonumber \\
t &=& - 2 m_e T_e, 
\end{eqnarray}
where $E_{\nu}$ and $T_e$ are the total energy of the incoming neutrino and the final kinetic energy of the electron. 
Eq. (\ref{2a}) then takes the familiar form 
\begin{equation}
\label{4}
\frac{d\sigma}{dT_e} = \frac{\alpha^2 \pi}{m_e^2} \mu_{\rm eff}^2 \left[ \frac{1}{T_e} - \frac{1}{E_\nu} \right] ,
\end{equation}
where $\mu_{\rm eff}$ is the effective neutrino magnetic moment measured at a distance $L$ from the neutrino source and written in units of Bohr magneton. It is given by 
\begin{equation}
\label{effmu}
\mu_{\rm eff}^2 = \sum_i \left| \sum_j U_{ej} e^{-iE_jL} \mu_{ji}\right|^2, 
\end{equation}
where the final neutrino states are summed over since they are not observed in the reactor experiments searching for the neutrino magnetic moment.  
Clearly at very short distances where the detectors of GEMMA and TEXONO detectors are placed, we can ignore the oscillating term if there are no sterile states. However, as we discuss later, possible contributions of the sterile states cannot be ignored.  

\begin{figure}
\begin{center}
\includegraphics[scale=0.2]{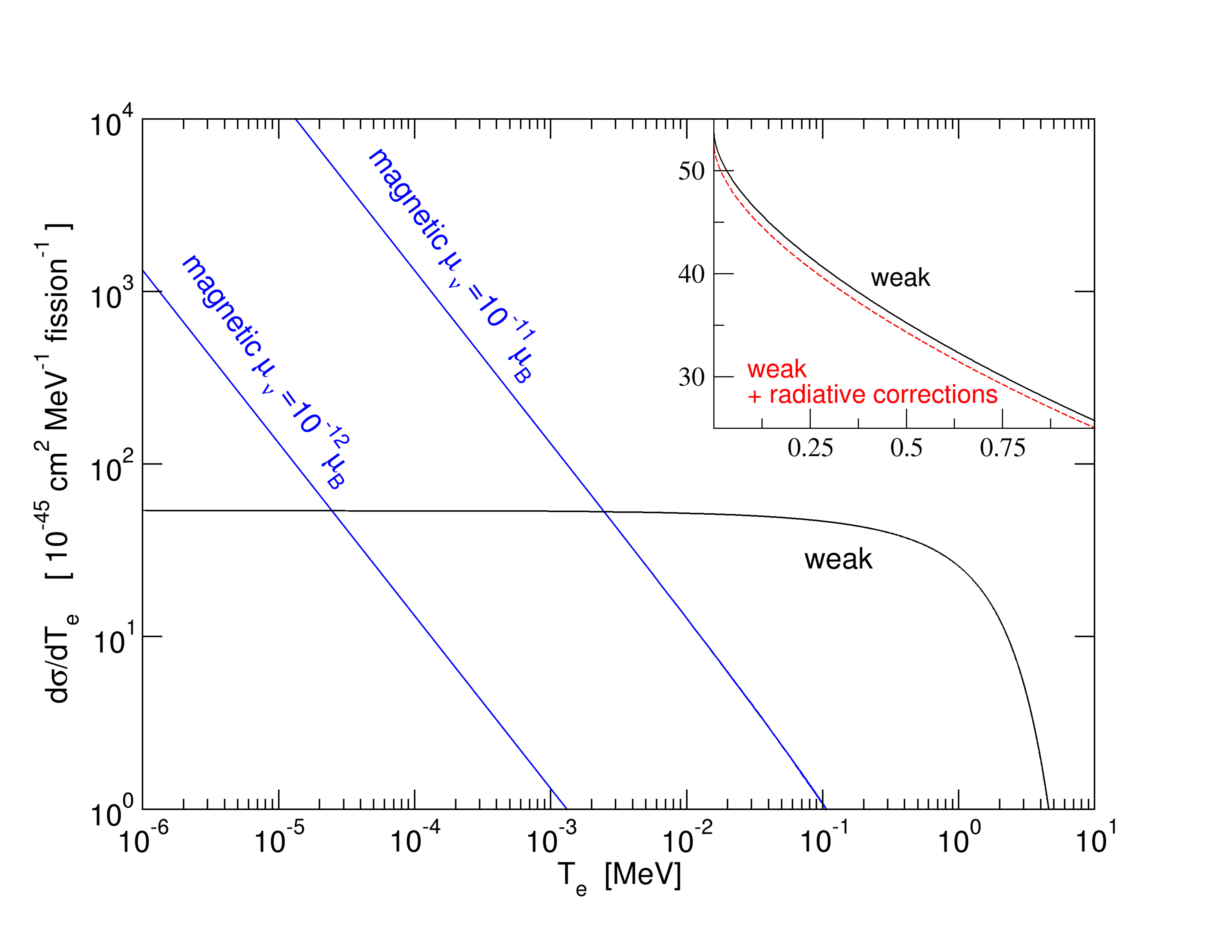}
\caption{(Color online) Log-log plot of the weak and electromagnetic components of the differential cross section for neutrino-electron scattering averaged over the antineutrino spectrum of fissioning $^{235}$U. The inset plot is the weak correction on the linear scale both with (dashed line) and without (solid line) radiative corrections \cite{Sarantakos:1982bp}.}
\label{fig:fig01a}
\end{center}
\end{figure}

Note that for massless neutrinos since the final states have different helicities in the magnetic moment and the standard weak scattering, they do not interfere. For massive neutrinos the interference term is proportional to the neutrino masses \cite{Grimus:1997aa} and we will ignore it.  The 
magnetic moment cross section will exceed the standard weak 
cross-section for recoil energies
\begin{equation}
\frac{T_e}{m_e} < \frac{\pi^2 \alpha^2}{G_F^2 (g_V^2 +g_A^2) m_e^4} \mu_{\nu}^2.  
\label{d2}
\end{equation}  
Hence reactor experiments aim to measure as low as possible electron recoil energies. To enable a comparison between the cross sections, we present the folded differential cross sections for the weak and electromagnetic components of neutrino-electron scattering in Fig. \ref{fig:fig01a}. The weak component of the cross section is plotted in the inset figure both with (dashed line) and without (solid line) radiative corrections. 

\section{Three flavors}

The Standard Model, minimally extended to include a massive neutrino, predicts a non-zero value of the magnetic moment. For Dirac neutrinos this prediction is \cite{Fujikawa:1980yx} 
\begin{equation}
\label{6}
\mu_{ij} = - \frac{e G_F}{8\sqrt{2}\pi^2} (m_i+m_j)  \sum_{\ell} U_{\ell i} U^*_{\ell j} f(r_{\ell}) 
\end{equation} 
with
\begin{equation}
f(r_{\ell}) \sim -\frac{3}{2} + \frac{3}{4} r_{\ell} + \cdots, \>\>\> r_{\ell}= \left( \frac{m_{\ell}}{M_W}\right)^2.  
\end{equation}
The electric dipole moment, $d_{ij}$,  of a Dirac neutrino is given by a similar expression to Eq. (\ref{6}), except that the term  $(m_i+m_j)$ 
is replaced by the term  $(m_i-m_j)$.   
For Majorana neutrinos only non-diagonal magnetic moments are permitted. In the case the CP-eigenvalues of the two neutrinos are opposite, 
the neutrino electric dipole moment is zero and the non-diagonal terms of the neutrino magnetic dipole moment are given by Eq. (\ref{6}), multiplied by a factor of 2 (see e.g. Ref. \cite{Mohapatra:1998rq}). 

\begin{figure}[t]
\begin{center}
\includegraphics[scale=0.163]{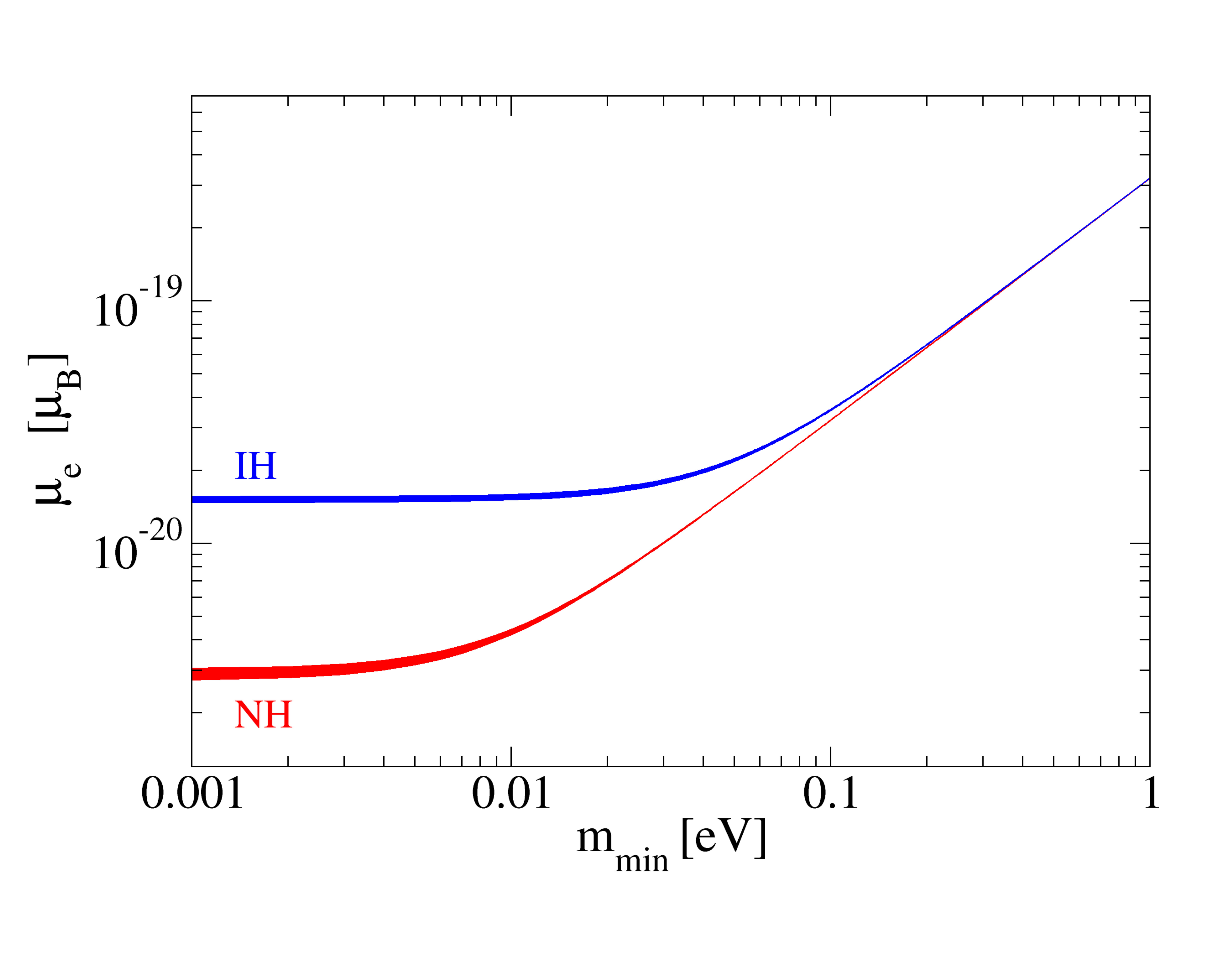}
\caption{(Color online) Standard Model prediction for the magnetic moment of a Dirac neutrino as function of the lowest neutrino mass, measured in a reactor experiment. NH and IH denote normal and inverted mass hierarchies, respectively. The neutrino mass differences and mixing angles are taken from the compilation of the Particle Data Group \cite{Beringer:1900zz}. The error bands represent the experimental errors in the mass splittings and the mixing angles.}
\label{fig:fig1}
\end{center}
\end{figure}

\begin{figure}[t]
\begin{center}
\includegraphics[scale=0.17]{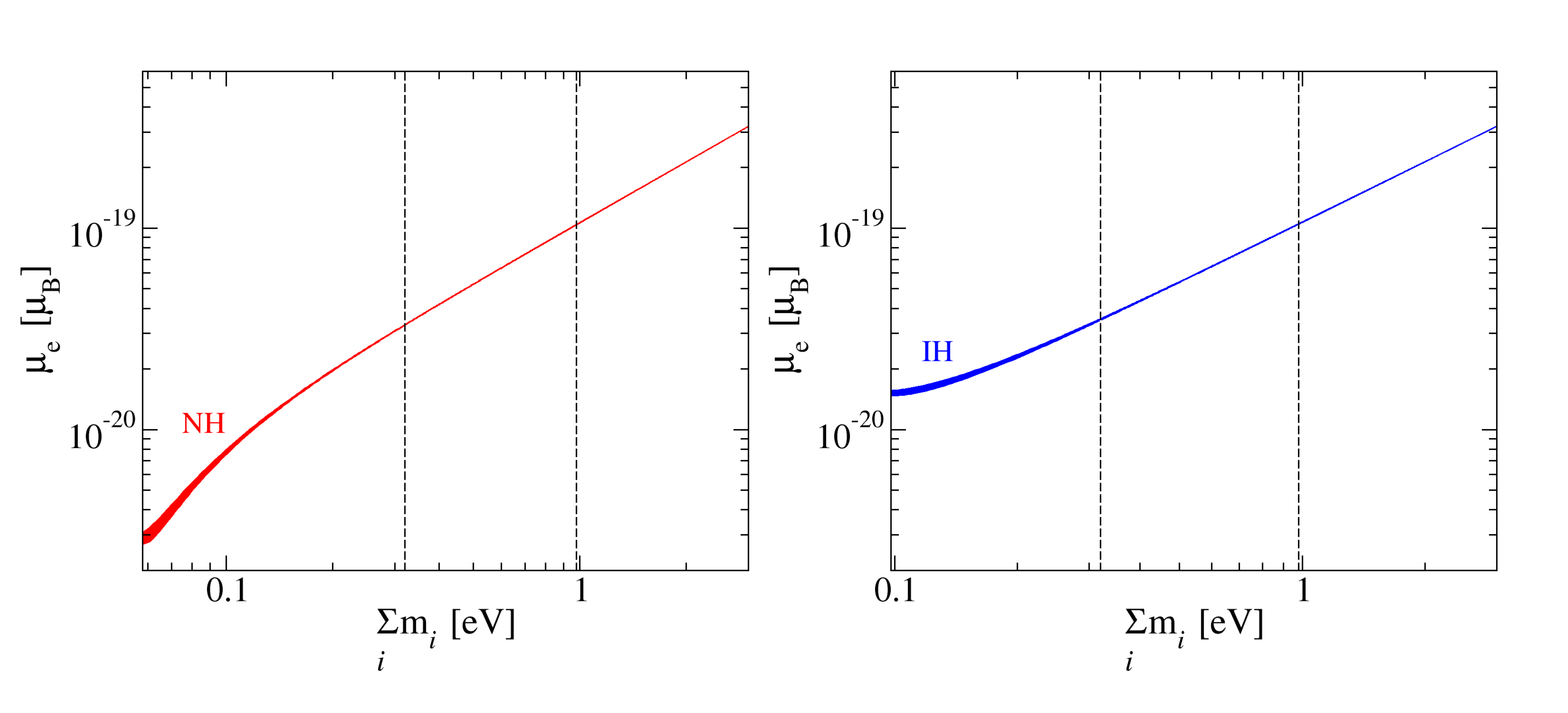}
\caption{(Color online) Standard Model prediction for the magnetic moment of a Dirac neutrino as a function of the sum of neutrino masses, measured in a reactor experiment. NH and IH denote normal and inverted mass hierarchies, respectively. The vertical lines denote the range  $0.32 \le \sum_i m_i \le 0.98$ eV from the analysis of the most recent Planck data \cite{Ade:2013zuv}.}
\label{fig:fig2}
\end{center}
\end{figure}

\begin{figure} 
\begin{center}
\includegraphics[scale=0.2]{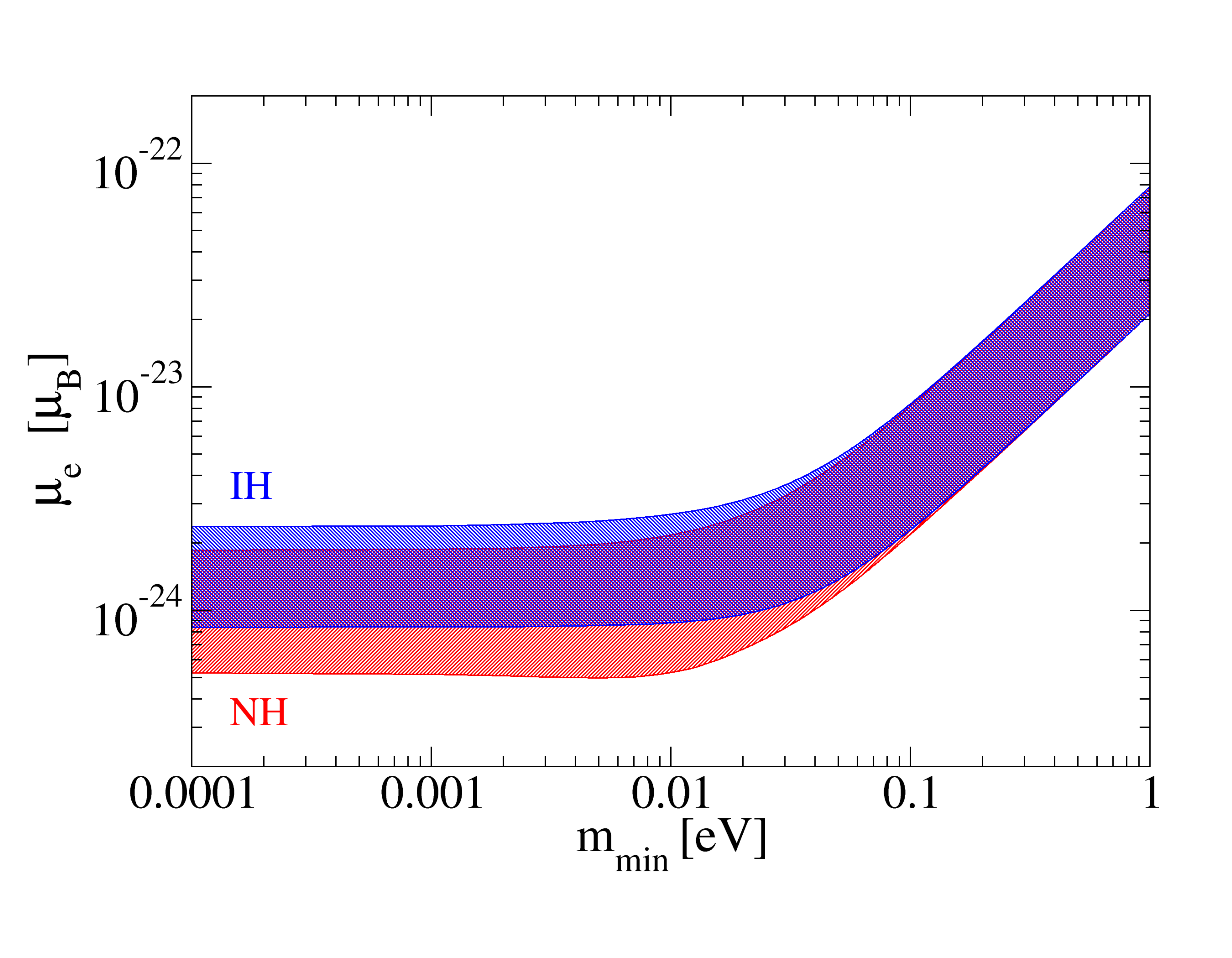}
\caption{(Color online) Standard Model prediction for the magnetic moment of a Majorana neutrino measured in a reactor experiment. NH and IH denote normal and inverted mass hierarchies, respectively. The error bands represent the experimental errors in the mass splittings and the mixing angles as well as the range of the Majorana phases $\alpha_1$ and $\alpha_2$ (see the Appendix)while keeping $\delta=0$.}
\label{fig:fig3}
\end{center}
\end{figure}

\begin{figure}
\begin{center}
\includegraphics[scale=0.2]{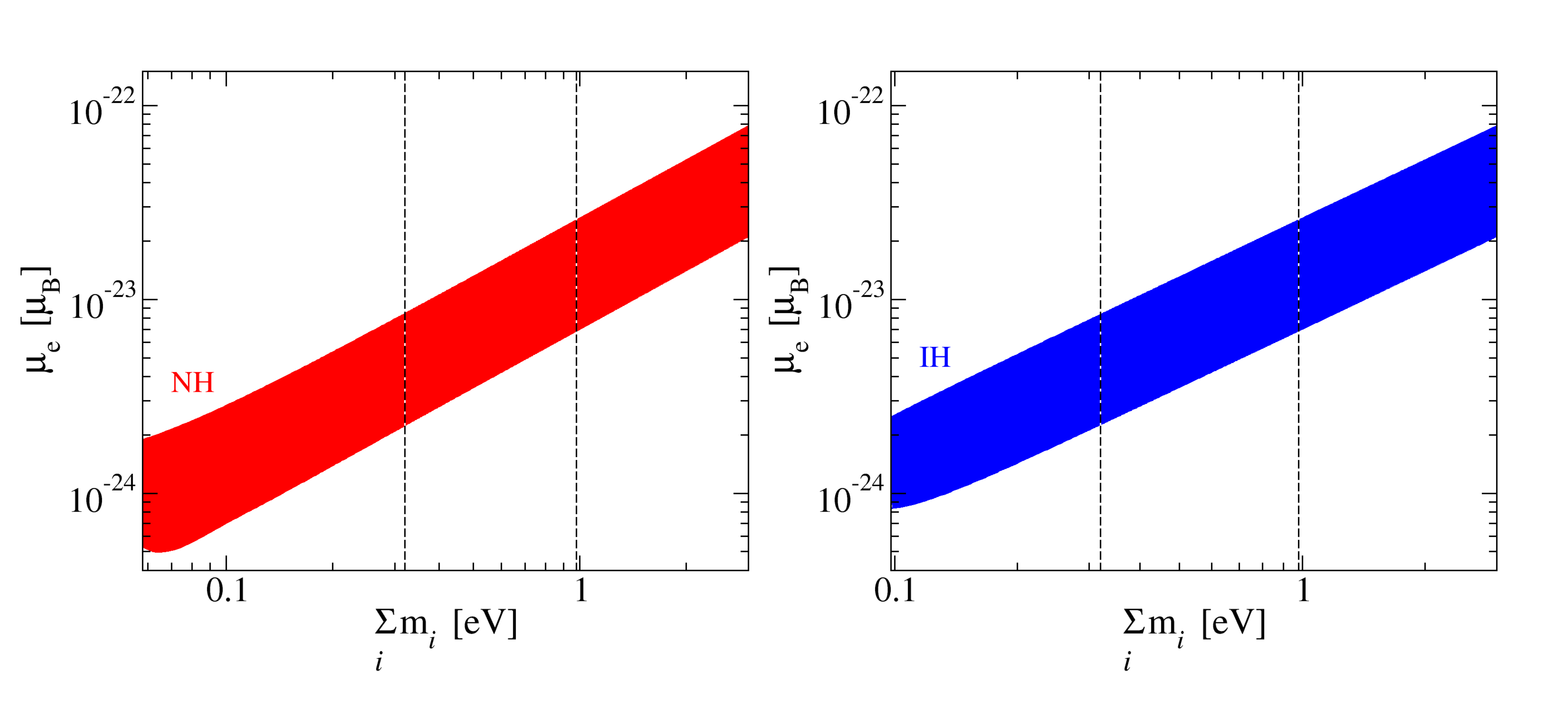}
\caption{(Color online) Same as Figure \ref{fig:fig2}, but for a Majorana neutrino. NH and IH denote normal and inverted mass hierarchies, respectively.}
\label{fig:fig4}
\end{center}
\end{figure}

Since the neutrino mass differences and all the mixing angles were recently measured with good accuracy, one can calculate the Standard Model prediction for the effective magnetic moment of Eq. (\ref{effmu}) as a function of the smallest neutrino mass. The result for the Dirac neutrinos is given in Fig. \ref{fig:fig1}.  
Clearly this magnetic moment is well below current experimental limits. 
It is possible to put an even tighter limit using cosmological arguments which limit the sum of all the neutrino masses \cite{Melchiorri:2006nj}. 
To illustrate this in Fig. \ref{fig:fig2} we plot the Standard Model prediction as a function of the sum of all the neutrino masses. 
Standard model predictions for the Majorana neutrino magnetic moment as a function of the lowest neutrino mass and the sum of neutrino masses are given in Figures \ref{fig:fig3} and \ref{fig:fig4}, respectively. These predictions are much lower than those for Dirac neutrinos. This is because in the Standard Model the diagonal contribution to the neutrino magnetic moment is dominant whereas the non-diagonal contributions are suppressed by a GIM-like mechanism (cf. Eq. \ref{6}). 

In the effective field theory approach, beyond the Standard Model physics is described by local operators at lower energies:
\begin{equation}
\label{effec}
{\cal L}= {\cal L}_{SM} + \frac{C^{(5)}}{\Lambda}  {\cal O}^{(5)} + \sum_i \frac{C^{(6)}_i}{\Lambda^2}  {\cal O}^{(6)}_i + \sum_i \frac{C^{(7)}_i}{\Lambda^3}  {\cal O}^{(7)}_i + \cdots ,
\end{equation}
where $\Lambda$ is the scale of the new physics, ${\cal O}^{(n)}_i$ is one of the mass dimension $n$ operators and $C^{(n)}_i$ is the associated multiplier, hoped to be order of unity. The unique dimension-five operator \cite{Weinberg:1979sa} is the Majorana neutrino mass and the dimension-seven operators include the Majorana magnetic moment. 

It is possible to give a general argument that would connect neutrino magnetic moment to the neutrino mass 
\cite{Balantekin:2006sw,Bell:2005kz,Bell:2006wi}. If the magnetic moment is generated by physics at scale $\Lambda$, we can generically write 
\begin{equation}
\label{mm}
\mu_{\nu} \sim \frac{e{\cal G}}{\Lambda}, 
\end{equation}
where ${\cal G}$ represents the combination of the coupling constants and other factors coming from the loop integrals that generate the magnetic moment. If we remove the external photon line from the diagrams leading to Eq. (\ref{mm}), we get a contribution to the neutrino mass of the order 
\begin{equation}
\delta m_{\nu} \sim {\cal G} \Lambda. 
\end{equation}
Hence in the case of Dirac neutrinos, magnetic moment terms induce radiative corrections to the neutrino mass of the order 
\begin{equation}
\label{mm1}
\delta m_{\nu} \sim \mu_{\nu} \Lambda^2. 
\end{equation}
A more careful analysis using effective field theories shows that the quadratic dependence on the energy scale arises from the quadratic divergence appearing in the renormalization of the dimension four neutrino mass operator \cite{Bell:2005kz}. Even for allowing a neutrino mass correction as large as $\sim 1$ eV (likely to be the fourth mass eigenstate), and for $\Lambda \sim 1$ TeV, the authors of Ref. \cite{Bell:2005kz} find a limit of $\mu_{\nu} \lesssim 8 \times 10^{-15} \mu_B$.  For the Majorana neutrinos, however, the limits are much weaker since mass matrix and magnetic moment matrix have different symmetries in flavor indices. For smaller values of $\Lambda$ ($\sim 1$ TeV), the best bound comes from the one loop contribution to the dimension seven operator in Eq. (\ref{effec}) \cite{Davidson:2005cs}, whereas for larger values of $\Lambda$ ($ \ge10$ TeV), the strongest bound comes from the two-loop contribution to the dimension five mass term 
\cite{Bell:2006wi}. Since the limits on Dirac neutrinos are more stringent, this suggests an observation of the neutrino magnetic moment of 
$\mu_{\nu} \gtrsim 10^{-15} \mu_B$ would indicate that neutrinos are Majorana particles. 

A summary of the recent status of the electromagnetic properties of neutrinos is given in Ref. \cite{Broggini:2012df} to which the reader is referred to for further details. Note that the neutrino magnetic moment is fundamentally defined in the mass basis. For Dirac neutrinos we have 
\begin{equation}
{\rm Dirac:}  \hskip 1cm \mu_{ij} = \mu^*_{ji} \Rightarrow \mu = \mu^{\dagger} .
\end{equation}
Neutrino magnetic moment can be written in the flavor basis as
\begin{equation}
\label{flavor}
\mu^{(F)}_{\alpha \beta} =  (U \mu U^{\dagger})_{\alpha \beta}, 
\end{equation}
where $U_{\alpha i}$ is the neutrino mixing matrix. It readily follows that, for Dirac neutrinos, the magnetic moment matrix is Hermitian in both mass and flavor bases. For Majorana neutrinos, diagonal terms are not permitted in the mass basis. In the mass basis the off-diagonal terms are imaginary and the magnetic moment matrix is antisymmetric \cite{Broggini:2012df} :
\begin{equation}
\label{maj1}
{\rm Majorana:} \hskip 1cm \mu_{ij} = - \mu_{ji}, \>\> \mu_{ij} = - \mu^*_{ij}.
\end{equation}
Recall that the arguments based on the effective field theory, summarized above, utilized the antisymmetry of the Majorana magnetic moment matrix in the flavor space. For three flavors, this is easiest to explore using the Gell-Mann SU(3) matrices, given in the Appendix. The neutrino mixing matrix of Eq. (\ref{matrix}) can be cast in the form Eq. (\ref{matrix2}). When all the phases are set equal to zero, this matrix is only acts in the SO(3) subspace. Clearly the Majorana magnetic moment matrix of Eq. (\ref{maj1}) is also an SO(3) matrix ($\lambda_2, \lambda_5$ and $\lambda_7$ are the only imaginary antisymmetric SU(3) matrices). Hence the transformation of Eq. (\ref{flavor}) leaves the magnetic moment matrix in the SO(3) subspace if all the phases are zero, i.e. Majorana magnetic moment matrix is also antisymmetric and purely imaginary in the flavor space. If the CP-violating phase $\delta$ is non-zero, then the Majorana magnetic moment matrix include diagonal terms in the flavor space proportional to $\sin \delta$ even if the Majorana phases $\alpha_1$ and $\alpha_2$ are ignored. 

Defining the matrix
\begin{equation}
\label{tdef}
T_{ij} = e^{-i E_jL} \delta_{ij}, 
\end{equation}
the effective neutrino magnetic moment $\mu_{\rm eff}^2$, measured at a distance $L$ from the reactor, can be written as 
\begin{equation}
\label{mueff2}
\mu_{\rm eff}^2 = (UT\mu \mu^{\dagger} T^{\dagger} U^{\dagger})_{ee}. 
\end{equation}
Note that in the same notation the electron neutrino survival amplitude, measured at a distance $L$ from the reactor, is
\begin{equation}
A(\nu_e \rightarrow \nu_e) = (U^{\dagger} T U)_{ee}. 
\end{equation}
The non-zero contributions of the T matrices to Eq. (\ref{mueff2})  come from neutrino mass differences. If there are no sterile neutrinos these are 
$ \delta m^2_{21}$ and $ \delta m^2_{31}$,
which are well known. Hence at distances where detectors are placed for neutrino magnetic moment measurements (few tens of meters), the T matrices can be replaced by the identity and we get
\begin{equation}
\label{mueff3}
\mu_{\rm eff}^2 = (\mu^{(F)} \mu^{(F)\dagger})_{ee}. 
\end{equation}

\section{Sterile neutrinos}

\begin{figure}
\begin{center}
\includegraphics[scale=0.16]{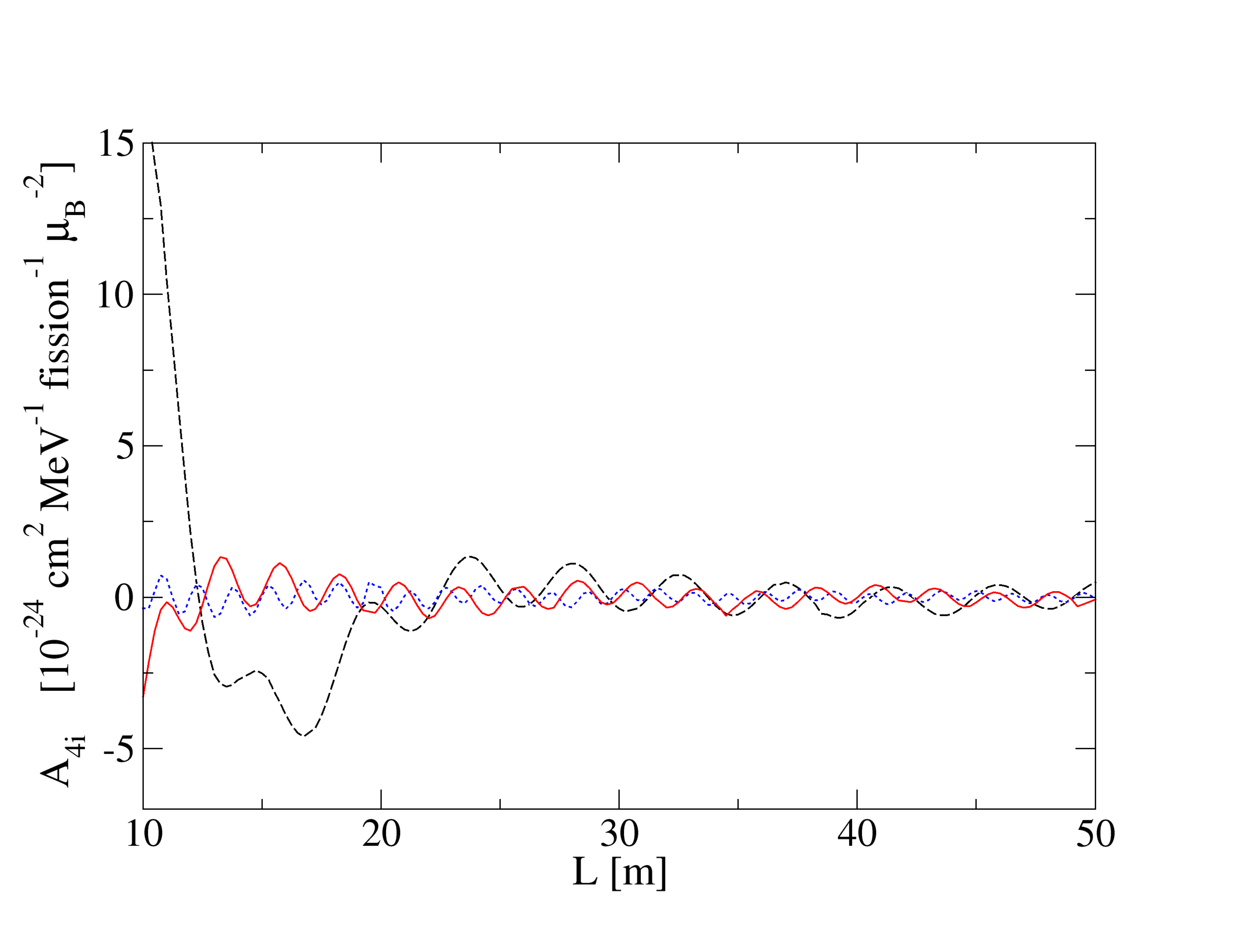}
\caption{(Color online) The contribution of the phases with arguments $(E_4-E_i) L \sim (\delta m_{4i}^2L)/2E$ to the count rate as a function of the distance from the point of neutrino production $L$. 
The dashed (black) line is for $\delta m_{4i}^2 = 1$ eV$^2$, the solid (red) line is for $\delta m_{4i}^2 = 1.78$ eV$^2$, and the
dotted (blue) line is for $\delta m_{4i}^2 = 3$ eV$^2$. The electron kinetic energy is taken to be $T_e = 2.8$ keV (GEMMA's threshold).}
\label{a4fig}
\end{center}
\end{figure}

If there are one or more sterile neutrinos, Eq. (\ref{mueff3}) no longer holds at distances where detectors are placed for neutrino magnetic moment measurements. For example for one sufficiently heavy sterile neutrino the phases with arguments $(E_4-E_i) L \sim (\delta m_{4i}^2L)/2E$ average to zero and we get 
\begin{equation}
\label{mu4}
\mu_{\rm eff}^2 = \sum_{i,j=1}^3 U_{ei} (\mu \mu^{\dagger})_{ij} U^{\dagger}_{je} + U_{e4} (\mu \mu^{\dagger})_{44} U^{\dagger}_{4e}. 
\end{equation} 
It is worthwhile to elaborate on the range of validity of this approximation. Note that the potential contribution from such terms to the count rate in the detectors can be seen from Eqs. (\ref{4}) and (\ref{effmu}) to be proportional to 
\begin{equation}
\label{a4}
A_{4i}=\int_{E_\nu,min}^{\infty}2 \frac{\alpha^2 \pi}{m_e^2}  \left[ \frac{1}{T_e} - \frac{1}{E_\nu} \right] \left[ \cos \left( \frac{\delta m^2_{4i}L}{2E_{\nu}}
\right) \right] \left( \frac{dN}{dE_{\nu}}\right) dE_{\nu}, 
\end{equation}
where $dN/dE_{\nu}$ is the incoming neutrino flux. We plot this contribution in Figure \ref{a4fig} as function of $L$ for several values of $\delta m^2$. 
In writing Eq. (\ref{a4}) we assumed that the terms multiplying $A_{4i}$ are real. If they were complex (e.g. with CP-violating phases) then a similar quantity with $\sin \left( \frac{\delta m^2_{4i}L}{2E_{\nu}} \right)$, which would behave in the same oscillatory manner, needs to be considered. 
Here $L$ is the distance between the point of production of the neutrino and the detector. An additional average over the detector core is not shown in this figure, but such an average would reduce the count rate further.  From this figure we see that for $\delta m_{4i}^2 \ge 1.78$ eV$^2$ (see below) the terms containing phases with arguments $(E_4-E_i) L \sim (\delta m_{4i}^2L)/2E$ in Eq. (\ref{mueff3}) average to zero. 

The survival probability of neutrinos at the same location can be written by averaging over the squares of the sines of the same phases as 
\begin{equation}
P( \bar{\nu}_e \rightarrow \bar{\nu}_e) = 1- 2 |U_{e4}|^2 + 2 |U_{e4}|^4. 
\end{equation}
In this work to be consistent we only use fits to the possible deficits to the reactor neutrino fluxes, since magnetic moments are also measured at the reactors. For one sterile neutrino, such a fit yields \cite{Kopp:2011qd} 
\begin{equation}
\label{e4}
|U_{e4}| \sim 0.151, \>\> \delta m_{41}^2 \sim 1.78 eV^2. 
\end{equation}

One should also note that non-diagonal components of the neutrino electric and magnetic dipole moments give rise to radiative decays of the mass eigenstates into one another, $\nu_i \rightarrow \nu_j + \gamma$, if the phase space permits it. This decay width in the rest frame is given by \cite{Marciano:1977wx,Petcov:1976ff,Pal:1981rm} 
\begin{equation}
\Gamma_{i \rightarrow j} = \frac{|\mu_{ij}|^2 + |\epsilon_{ij}|^2}{8\pi} \left( \frac{m_i^2 - m_j^2}{m_i} \right)^3 
= 5.308 s^{-1} \left( \frac{ \mu_{{\rm total},ij}}{\mu_B} \right)^2 \left( \frac{m_i^2 - m_j^2}{m_i^2} \right)^3 \left( \frac{m_i}{{\rm eV}}\right)^3, 
\end{equation}
where $\mu_{{\rm total},ij}^2$ includes contributions from both electric and magnetic moments of the neutrino. Neglecting the active neutrino masses, this yields a sterile state lifetime of
\begin{equation}
\tau_{4 \rightarrow i} \sim \left( \frac{\mu_B}{\mu_{4i}}\right)^2 \times 10^{-9} y , \> i=1,2,3. 
\end{equation}
Even for the GEMMA upper limit on the neutrino magnetic moment, this gives a lifetime of longer than $1.18 \times 10^{12}$ years,  significantly longer than the age of the Universe, $13.8 \times 10^{9}$ years. Hence such a sterile state can treated as stable. 

Eq. (\ref{mu4}) can be regrouped as
\begin{equation}
\label{zz1}
\mu_{\rm eff}^2 = \sum_{k=1}^3 \left| \sum_{i=1}^3 U_{ei} \mu_{ik} \right|^2  +  \left| \sum_{i=1}^3 U_{ei} \mu_{i4}  \right|^2 + \left| U_{e4} \right|^2 \sum_{i=1}^3 \mu_{i4}\mu_{4i} , 
\end{equation}
where we have assumed that neutrinos are Majorana particles. Applying the Cauchy-Schwarz inequality to the individual sums in Eq. (\ref{zz1}), we obtain the inequality 
\begin{equation}
\label{zzz1}
\mu_{\rm eff}^2 \le \sum_{i=1}^3 \mu_{i4}^2 + \left(1- \left| U_{e4} \right|^2 \right) \sum_{i,j=1}^3 \mu_{ij}^2. 
\end{equation}
\begin{figure}
\begin{center}
\includegraphics[scale=0.16]{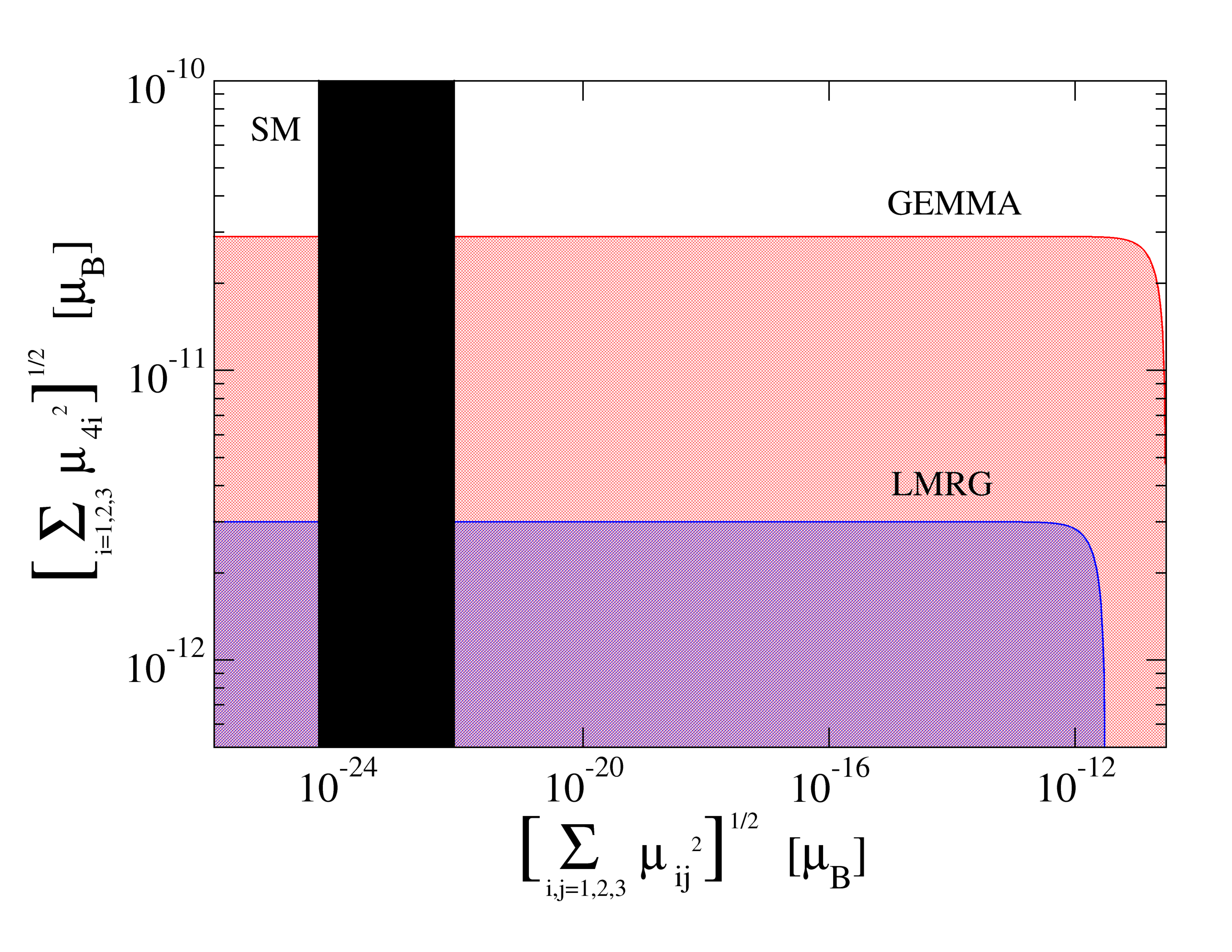}
\caption{(Color online) The allowed regions of the sterile neutrino contribution $\mu_{i4}^2$ versus the combination $\sum_{i,j=1}^3 \mu_{ij}^2$ coming from the active flavors only in the case that neutrinos are Majorana. The GEMMA and the low-mass red giant (LMRG) limits are shown. The black band is the standard model prediction for the active flavor masses between $10^{-4}$ and $1$ eV.}
\label{fig:fig6}
\end{center}
\end{figure}
In Figure \ref{fig:fig6} we plot the allowed regions of the sterile neutrino contribution $\mu_{i4}^2$ versus the combination $\sum_{i,j=1}^3 \mu_{ij}^2$ coming from the active flavors only. The two allowed regions shown correspond to the GEMMA's upper bound of $\mu_{\nu} \le 2.9 \times 10^{-11} \mu_B$ and the red giant limit of $\mu_{\nu} \le 3 \times 10^{-12} \mu_B$. For comparison the Standard Model prediction with active flavors with masses between $10^{-4}$ and $1$ eV is also shown as a black band.

\section{Conclusions}

Finding a more conclusive value for the neutrino magnetic moment is clearly an important goal as it not only affects astrophysical processes, but reveals much needed information about the nature of the neutrino. As we have shown, given that most of the neutrino parameters have been well measured, the prediction of the standard model will be pretty much pinned down once the absolute value of the neutrino mass and hierarchy are also known. Within the minimum neutrino mass range of $10^{-4}$ to $1$ eV, the ranges for the electron neutrino magnetic moment given by the standard model for both Dirac [$3\times 10^{-21}$, $3\times 10^{-19}$] $\mu_{B}$ and Majorana [$5\times 10^{-25}$, $8\times 10^{-23}$] $\mu_{B}$ are currently well below experimental observation. The recoil energies reactor based experiments would need to be able to detect to confirm standard model predictions are impractically small (a neutrino magnetic moment on the order of $10^{-19}\mu_{B}$ would require the ability to observe an electron recoil energy on the order of $10^{-12}$ to $10^{-13}$ eV for the cross section to be visible over the standard weak background). However, pushing experimental thresholds lower is still desirable since it is evident that the effective magnetic moment is greatly influenced by physics beyond the standard model.

The presence of physics beyond the standard model in the form of a light sterile neutrino and whether neutrinos are Dirac or Majorana particles are both areas of active research which would be directly affected by the observation of a neutrino magnetic moment that is larger than what is predicted by the standard model. Reactor experiments, such as GEMMA and TEXONO, working to measure this quantity give the most reliable upper bounds since the values given by astrophysical considerations, such as low-mass red giant cooling, are highly model dependent. Here we have illustrated that a light sterile neutrino could contribute to the effective magnetic moment observed by electron-neutrino scattering experiments, but would not affect the kinematics. If future reactor experiments are able to confirm the existence of a light sterile neutrino, the observation of a large effective magnetic moment would suggest at least two possibilities. One possibility is that the three known active neutrinos have additional interactions which significantly contribute to their magnetic moments. The other possibility is that although sterile neutrinos do not interact weakly, their magnetic moments are much larger than those of the three standard active neutrinos. Such interactions would greatly impact the physics that goes into examinations of astrophysical processes such as core-collapse supernovae.

This work was supported in part 
by the U.S. National Science Foundation Grant No.  PHY-1205024, in part by the
University of Wisconsin Research Committee with funds
granted by the Wisconsin Alumni Research Foundation. 

\appendix

\section{Properties of the Fundamental Representations of SU(3) and SU(4)}

We take the eight Gell-Mann matrices for SU(3) to be 
\begin{eqnarray}
\lambda_1 &=& \left(   
   \begin{matrix} 
      0 & 1 & 0 \\
      1 & 0 & 0 \\
      0 & 0 & 0 \\
   \end{matrix} \right),
\lambda_2 = \left(   
   \begin{matrix} 
      0 & -i & 0 \\
      i & 0 & 0 \\
      0 & 0 & 0 \\
   \end{matrix} \right),
   \lambda_3 = \left(   
   \begin{matrix} 
      1 & 0& 0 \\
      0 & -1 & 0 \\
      0 & 0 & 0 \\
   \end{matrix} \right),
\lambda_4 = \left(   
   \begin{matrix} 
      0 & 0 & 1 \\
      0 & 0 & 0 \\
      1 & 0 & 0 \\
   \end{matrix} \right), \nonumber \\
   \lambda_5 &=& \left(   
   \begin{matrix} 
      0 & 0 & -i \\
      0 & 0 & 0 \\
      i & 0 & 0 \\
   \end{matrix} \right),
\lambda_6 = \left(   
   \begin{matrix} 
      0 & 0 & 0 \\
      0 & 0 & 1 \\
      0 & 1 & 0 \\
   \end{matrix} \right),
   \lambda_7 = \left(   
   \begin{matrix} 
      0 & 0& 0 \\
      0 & 0 & -i \\
      0 & i & 0 \\
   \end{matrix} \right),
\lambda_8 =  \frac{1}{\sqrt{3}}\left(   
   \begin{matrix} 
      1 & 0 & 0 \\
      0 & 1 & 0 \\
      0 & 0 & -2 \\
   \end{matrix} \right).
\end{eqnarray}
The $3 \times 3$ neutrino mixing matrix,  
\begin{equation}
\label{matrix}
U = \left(
\begin{array}{ccc}
 1 & 0  & 0  \\
  0 & C_{23}   & S_{23}  \\
 0 & -S_{23}  & C_{23}  
\end{array}
\right)
\left(
\begin{array}{ccc}
 C_{13} & 0  & S_{13} e^{-i\delta_{CP}}  \\
 0 & 1  & 0  \\
 - S_{13} e^{i \delta_{CP}} & 0  & C_{13}  
\end{array}
\right) 
\left(
\begin{array}{ccc}
 C_{12} & S_{12}  & 0  \\
 - S_{12} & C_{12}  & 0  \\
0  & 0  & 1  
\end{array}
\right)
\left(
\begin{array}{ccc}
 1 & 0  & 0  \\
 0 & e^{i\alpha_1/2}  & 0  \\
0  & 0  &   e^{i\alpha_2/2}
\end{array}
\right), 
\end{equation}
where $C_{ij} = \cos \theta_{ij}$, $S_{ij} = \sin \theta_{ij}$, $\delta_{CP}$ is the CP-violating phase and $\alpha_{1,2}$ are the Majorana phases, can be written in terms of the Gell-Mann matrices as
\begin{equation}
\label{matrix2}
U = e^{i\theta_{23} \lambda_7} e^{-i \delta (\lambda_3+\sqrt{3} \lambda_8)/4} e^{i\theta_{13} \lambda_5} 
e^{+i \delta (\lambda_3+\sqrt{3} \lambda_8)/4} e^{i\theta_{12} \lambda_2} e^{i\left[\frac{\alpha_1+\alpha_2}{6} I + \frac{\alpha_1-2\alpha_2}{4\sqrt{3}}\lambda_8 + \frac{\alpha_1}{4} \lambda_3 \right]} .
\end{equation}
It can easily be verified that $\lambda_7, \lambda_5$, and $\lambda_2$ form the SO(3) subalgebra of SU(3). Hence if all the phases are set equal to zero, then the neutrino mixing matrix becomes an SO(3) transformation. 

For four flavors we need to consider the fundamental representation of the SU(4) group. We will only write down the six matrices which are antisymmetric. Three of those are $\lambda_7, \lambda_5$, and $\lambda_2$ of SU(3), embedded into four by four matrices:
\begin{equation}
L_1 = \left(   
   \begin{matrix} 
      0 & 0& 0 &0 \\
      0 & 0 & -i &0\\
      0 & i & 0 &0\\
      0 & 0 & 0 &0 \\
   \end{matrix} \right),
 L_2=  \left(   
   \begin{matrix} 
      0 & 0 & -i & 0\\
      0 & 0 & 0 & 0\\
      i & 0 & 0 & 0\\
      0 & 0 & 0 & 0 \\
   \end{matrix} \right),  
 L_3 =  \left(   
   \begin{matrix} 
      0 & -i & 0 & 0\\
      i & 0 & 0 & 0\\
      0 & 0 & 0 & 0\\
      0 & 0 & 0 & 0\\
   \end{matrix} \right). 
\end{equation}
The other three include the fourth mass eigenstate:
\begin{equation}
K_1 = \left(   
   \begin{matrix} 
      0 & 0& 0 & -i \\
      0 & 0 & 0 &0\\
      0 & 0 & 0 &0\\
      i & 0 & 0 &0 \\
   \end{matrix} \right),
 K_2=  \left(   
   \begin{matrix} 
      0 & 0 & 0 & 0\\
      0 & 0 & 0 & -i\\
      0 & 0 & 0 & 0\\
      0 & i & 0 & 0 \\
   \end{matrix} \right),  
 K_3 =  \left(   
   \begin{matrix} 
      0 & 0 & 0 & 0\\
      0 & 0 & 0 & 0\\
      0 & 0 & 0 & -i\\
      0 & 0 & i & 0\\
   \end{matrix} \right). 
\end{equation}
Again it is easy to verify that 
\begin{equation}
\frac{1}{2} (L_i + K_i ), \>\>  i=1, 2, 3,
\end{equation}
and 
\begin{equation}
\frac{1}{2} (L_i - K_i ), \>\>  i=1, 2, 3,
\end{equation}
generate two mutually commuting SO(3) algebras. Hence, if all the phases are set equal to zero, the four-flavor neutrino mixing matrix becomes an  
$ SO(4) \sim SO(3) \times SO(3)$ transformation.

\end{document}